\documentclass[10pt]{IEEEtran}
\usepackage[dvips]{graphicx}
\newcommand{\qed}{\fbox{}}
\newcommand{\ve}[1]{ \mbox{\boldmath$#1$} }
\newcommand{\defeq}{\stackrel{\triangle}{=}}
\newtheorem{example}{Example}
\newtheorem{definition}{Definition}

\newtheorem{lemma}{Lemma}
\newtheorem{corollary}{Corollary}
\newtheorem{theorem}{Theorem}

\newcommand{\Aen}{{\cal A}}
\newcommand{\Ben}{{\cal B}}
\newcommand{\Gen}{{\cal G}}
\newcommand{\Een}{{\cal E}}
\newcommand{\Ren}{{\cal R}}
\newcommand{\Cen}{{\cal C}}
\newcommand{\Fen}{{\cal F}}

\title{Average Stopping Set Weight Distribution of Redundant Random Matrix Ensembles}
\author{Tadashi Wadayama\thanks{Nagoya Institute of Technology, Gokiso-cho, Showa-ku, Nagoya, Aichi, Japan, 
email: wadayama@nitech.ac.jp}}
 
\begin{document}
\maketitle

\section*{Abstract}
In this paper, redundant random matrix ensembles (abbreviated as {\em redundant random ensembles})
are defined and their stopping set (SS) weight distributions are analyzed. 
A redundant random ensemble consists of a set of binary matrices with linearly 
dependent rows.
These linearly dependent rows (redundant rows) significantly reduce
the number of stopping sets of small size.
An upper and lower bound on the average SS weight distribution of 
the redundant random ensembles are shown. 
From these bounds, the trade-off 
between the number of redundant rows (corresponding to decoding complexity 
of BP on BEC) and the critical exponent of the asymptotic growth rate of SS weight distribution
(corresponding to decoding performance) can be derived.
It is shown that, in some cases,
a dense matrix with linearly dependent rows yields asymptotically 
({\em i.e.}, in the regime of small erasure probability) better performance
than regular LDPC matrices with comparable parameters.

\subsection*{Keywords}
LDPC codes, Stopping set, Weight distribution, Ensemble

\section{Introduction}
On binary erasure channel (BEC), 
the decoding performance of belief propagation (BP)-based iterative decoder 
of low-density parity-check(LDPC) codes
is dominated by 
 combinatorial structures in a Tanner graph, which are called {\em stopping sets} (SS)\cite{Di2002}.
Di et al.\cite{Di2002} introduced the idea of stopping sets and 
presented a recursive method to evaluate
the average block and bit error probabilities of LDPC codes\cite{Gal63} of finite  length on BEC\cite{Di2002}.
Orlitsky et al. \cite{Orlitsky} found the asymptotic behavior of the SS weight
distributions of bipartite graph ensembles and extended  the results of Di et al. to the irregular code case.

For a given binary linear code $C$, 
it is hoped to find 
the best representation of $C$ ({\em i.e.}, a parity check matrix) which yields
the smallest block (or bit) error probability when it is decoded with 
iterative decoding on BEC. A parity check matrix which defines $C$ can be 
a {\em redundant parity check matrix}, which is not a full-rank matrix:
that is, it can contain some linearly dependent rows. For example,
some finite geometry LDPC codes require a redundant parity check matrix 
to achieve  good decoding performance with BP.
Recent works of
Schwartz and Vardy\cite{Schwartz}, Abdel-Ghaffar and Weber\cite{ghaffar},
Hollmann and Tolhuizen\cite{Hollmann} indicate that the stopping set weight 
distribution of a given matrix can be improved by appending linearly dependent rows to the
original matrix.

Recent developments described in studies of the average weight 
distributions of LDPC codes, such as Litsyn and Shevelev\cite{LS02}\cite{LS03}, Burshtein and Miller\cite{MV04}
Richardson and Urbanke\cite{modern},
imply that {\em ensemble analysis} is a powerful method for investigating
typical properties of codes and matrices, properties are not easy to obtain from an instance.
Furthermore, from the asymptotic behavior of typical properties such as these, 
we often can  predict a threshold phenomenon.

The average stopping set weight distributions presented in \cite{Di2002} and \cite{Orlitsky}
are a useful decoding performance measure (for BP on BEC) 
of a given ensemble of parity check matrices.
The distribution can be used for optimizing an ensemble suitable for BEC. 
BEC is not only of practical interest, but also can be considered as a 
good starting point for theoretical studies of performance analysis of BP for more general
channels, such as binary input  symmetric output channels\cite{modern}.

In this paper, redundant random matrix ensembles (abbreviated as {\em redundant random ensembles})
are defined and their SS weight distributions are analyzed. 
The redundant random ensemble consists of a set of binary matrices with linearly 
dependent rows.
These linearly dependent rows (redundant rows) significantly reduce 
the number of stopping sets of small size.
An upper bound and a lower bound on the average SS weight distribution of 
redundant random ensemble will be shown. 
From these bounds, the trade-off 
between the number of redundant rows (corresponding to decoding complexity 
of BP) and the critical exponent of the asymptotic growth rate of SS weight distribution
(corresponding to decoding performance) can be derived.

\section{Average SS weight distribution}
In this section, some notation and definitions required in the paper are introduced.
Furthermore, some known results on average SS weight distributions are briefly reviewed.

\subsection{Stopping set and SS weight distribution}
Let $F_2$ be the binary Galois field with elements $\{0,1\}$.
The operator $\circ$ denotes the integer ring inner product defined by
$
\ve h \circ \ve x \defeq h_1 x_1 + h_2 x_2 + \cdots + h_n x_n
$
for $\ve h \defeq (h_1,h_2,\ldots,h_N) \in  F_2^n$ and
$\ve x \defeq(x_1,x_2,\ldots,x_n) \in  F_2^n$ $(n \ge 1 )$. The additions in the
above definition of $\circ$ is the addition of the integer ring ({\em i.e.}, $1+1=2$).
In this paper, the addition of $F_2$ is denoted by $\oplus$ ({\em i.e.}, $1\oplus 1=0$).


For a given $\ve x \in F_2^n$ and an $m \times n$ binary matrix $H$ $(m, n \ge 1)$, 
the {\em SS indicator} $q_H( \ve x)$ is defined by
\begin{equation}
q_H(\ve x) \defeq \# \{i \in [1,m]:  \ve h_i \circ \ve x = 1  \},
\end{equation}
where $\ve h_{i}$ denotes the $i$-th row vector of $H$, and we denote the cardinality of a given
finite set $X$ by $\# X$.
The notation $[a,b]$ means the set of consecutive integers from $a$ to $b$.
The stopping set is defined as follows:
\begin{definition}[Stopping set]
If $q_H(\ve x)=0$ then
$\ve x \in F_2^n$ is called a {\em SS vector} of $H$.
The support set of $\ve x$, 
\begin{equation}
S_{\ve x} \defeq \{i \in [1,n]: x_i = 1 \},
\end{equation}
is called a stopping set of $H$\footnote{This definition of SS is not exactly the same as the original definition\cite{Di2002}. The present definition 
covers the case where there exists a variable node without an edge({\em i.e.}, a zero column). }.
\hfill\qed
\end{definition}
Note that  if there exists a row vector $\ve h_i (i \in [1,m])$  satisfying 
$\ve h_i \circ \ve x = 1$ then $\ve x$ is not an SS vector. 
Let $\ve y \defeq (y_1,y_2, \ldots, y_n) \in \{0,1,e \}^n$ be a received word through a BEC,
where $e$ denotes the erasure symbol. It is known that BP fails to decode $\ve y$
if and only if the erasure support set $E_{\ve y} \defeq \{i \in [1,n]: y_i = e \}$ contains a non-empty stopping set of $H$. 
This property justifies the study  of SSs in order to reveal 
the BP decoding performance for BEC.

The next definition provides the definition of the SS weight distribution and
the stopping distance:
\begin{definition}[SS weight distribution and stopping distance]
For a given $m \times n (m, n \ge 1)$ matrix $H$, the SS weight distribution $\{S_w(H) \}_{w=0}^n$
is defined by
\begin{equation}
S_w(H) \defeq	 \sum_{\ve x \in Z^{(n,w)}} I[q_H(\ve x) = 0]
\end{equation}
for $0 \le w  \le n$,  where $Z^{(n,w)}$ is the set of constant weight binary vectors of length $n$ whose
Hamming weights are $w$. The notation $I[condition]$ is the indicator function
such that $I[condition] = 1$ if $condition$ is true; otherwise, it gives 0.
The stopping distance of $H$ is defined by
\begin{equation}
\Delta(H) \defeq \min\{w \in [1,n]: S_w(H) \ne 0 \}.
\end{equation}
\hfill\qed
\end{definition}
\begin{example}
Let 
\begin{equation}
H = \left(
\begin{array}{ccc}
1 & 0 & 1 \\
1 & 1 & 1 \\
\end{array}
\right).
\end{equation}
In this case, $W \defeq \{\emptyset, \{1,3\}, \{1,2,3\} \}$ is the set of stopping sets of $H$ and
$2^{[1,n]} \backslash W$ is the set of non-stopping sets of $H$. The SS weight distribution is
given by $\{S_w(H) \}_{w=0}^3 = \{1,0,1,1\}$ and the stopping distance is $\Delta(H) = 2$.
\hfill\qed
\end{example}

\subsection{Average SS weight distribution}

Suppose that $\Gen$ is a set of binary $m\times n$ matrices$(m, n \ge 1)$.
Note that we allow the possibility that  $\Gen$ may contain some matrices with the same 
configuration. Such matrices should be distinguished as distinct matrices. 
We assign the same probability, $1/\#\Gen$, to each matrix in $\Gen$.
Let $f(H)$ be a real-valued function which depends on $H \in \Gen$.
The expectation  of $f(H)$ with respect to the ensemble $\Gen$ is defined by
\begin{equation}
E_{\Gen}[f(H)] \defeq \sum_{H \in \Gen} P(H) f(H) = \frac{1}{\# \Gen} \sum_{H \in \Gen} f(H).
\end{equation}
The average SS weight distribution is defined as follows:
\begin{definition}[Average SS weight distribution]
The average SS weight distribution of a given ensemble $\Gen$ is defined by
\begin{equation}
S_w^{\Gen} \defeq E_\Gen[S_w(H) ]
\end{equation}
for $0 \le w \le n$.
\hfill\qed
\end{definition}
One of the most important properties of an ensemble is 
its symmetry. Although several types of symmetry are shown to be useful in \cite{wadayama0},
the following simple definition is sufficient for the purposes of this paper.
\begin{definition}[Symmetry of an ensemble]
If the equality
\begin{equation}
\# \{H \in \Gen:  q_H(\ve x_1) = 0 \} = \# \{H \in \Gen:  q_H(\ve x_2) = 0 \}
\end{equation}
holds for any $\ve x_1, \ve x_2 \in Z^{(n,w)}$ and any $w \in [0,n]$, then
the ensemble $\Gen$ is called {\em symmetric}.
\hfill\qed
\end{definition}
There is a simple expression of the average SS weight distribution for a symmetric ensemble.
\footnote{Note that all the ensembles discussed in this paper are symmetric.}.
The next lemma shows that the evaluation of the average SS weight distribution is
equivalent to a counting problem of matrices satisfying a certain condition.
\begin{lemma}\label{synen}
If $\Gen$ is symmetric, then
\begin{equation}
S_w^{\Gen}  = \frac{{n \choose w}}{\# \Gen} \# \{H \in \Gen:  q_H(\ve x_w) = 0 \}
\end{equation}
holds for $w \in [0,n]$. The vector $\ve x_w \in Z^{(n,w)}$ is 
the binary vector whose first $w$-elements are one, with all other elements zero.
 \\
(Proof)
The average SS weight distribution of  $\Gen$ can be transformed into the following form,
\begin{eqnarray} \nonumber 
S_w^{\Gen} &=& E_\Gen[S_w(H) ]\\ \nonumber
&=&\sum_{H \in \Gen} P(H) \sum_{\ve x \in Z^{(n,w)}} I[q_H(\ve x) = 0] \\ \nonumber
&=&\sum_{\ve x \in Z^{(n,w)}} \sum_{H \in \Gen} P(H)  I[q_H(\ve x) = 0] \\
&=&\frac{1}{\# \Gen}\sum_{\ve x \in Z^{(n,w)}} \# \{H \in \Gen:  q_H(\ve x) = 0 \} \\
&=& \frac{{n \choose w}}{\# \Gen} \# \{H \in \Gen:  q_H(\ve x_w) = 0 \}.
\end{eqnarray}
The last equation follows from the assumption;
namely $\# \{H \in \Gen:  q_H(\ve x) = 0 \}$ takes the same value for any $\ve x \in Z^{(n,w)}$. 
\hfill\qed
\end{lemma}

\subsection{Average SS weight distributions of known ensembles}
In this subsection, the average SS weight distribution of three well-known ensembles:
{\em the random ensemble, the constant row weight ensemble and the bipartite ensemble}, 
will be shown.

\subsubsection{Random ensemble}
The random ensemble $\Ren_{m,n}$ is the set of all binary $m \times n$ matrices$(m, n \ge 1)$.
Thus, the size of $\Ren_{m,n}$ is equal to $2^{mn}$.
The following lemma gives the average SS distribution of the random ensemble.
The key of the proof is to count $\# \{H \in \Ren_{m,n}:  q_H(\ve x_w) = 0 \}$.

\begin{lemma}
\label{AveSSrand}
The average SS distribution of the random ensemble $\Ren_{m,n}$ is given by 
\begin{equation}\label{nonextend}
S_w^{\Ren_{m,n}}  = {n \choose w} (1 - w 2^{-w})^m
\end{equation}
for $0 \le w \le n$.
\\
(Proof)
From the definition of the ensemble, it is evident that the following equality holds,
\begin{equation}
\# \{H \in\Ren_{m,n}:  q_H(\ve x_w) = 0 \} = (\# \{\ve h \in F_2^n: \ve h \circ \ve x_w \ne 1 \})^m.
\end{equation}
It is easy to show that the equality
\begin{equation}
\# \{\ve h \in F_2^n: \ve h \circ \ve x_w \ne 1 \} = (2^{w}-w) 2^{n-w}
\end{equation}
holds, since
\begin{equation}
\# \{\ve h \in F_2^n: \ve h \circ \ve x_w = 1 \} = w 2^{n-w}.
\end{equation}
Combining the above results, we get
\begin{eqnarray} \nonumber
S_w^{\Ren_{m,n}} \hspace{-3mm}&=& \hspace{-2mm}E_{\Ren_{m,n}}[S_w(H) ]\\ \nonumber
&=&\hspace{-2mm}\frac{1}{\# \Ren_{m,n}}
{n \choose w} \# \{H \in \Ren_{m,n}:  q_H(\ve x_w) = 0 \} \\ \nonumber
&=&\hspace{-2mm}\frac{1}{2^{mn}}{n \choose w} (2^n - w 2^{n-w})^m \\
&=& \hspace{-2mm}{n \choose w} (1 - w 2^{-w})^m.
\end{eqnarray}
Note that in deriving the second equality from the first equality, the symmetric property of the random ensemble and Lemma \ref{synen} was used.
\hfill\qed
\end{lemma}

The {\em asymptotic growth rate} of the average SS distribution (for simplicity, abbreviated as the asymptotic growth rate) 
of the random ensemble  is defined by
\begin{equation} \label{asymptoSSdef}
\lambda_\ell \defeq \lim_{n \rightarrow \infty} \frac 1 n \log_2 S_{\ell n}^{\Ren_{(1-R)n,n}},
\end{equation}
where $R (0 < R < 1)$ is called {\em design rate} and $\ell (0 \le \ell \le 1)$ is
the {\em normalized weight}.
The asymptotic growth rate reflects the asymptotic (in the limit as $n$ goes to infinity) behavior of the average SS 
weight distribution for fixed design rate and normalized weight.
The next lemma gives the asymptotic growth rate of random ensembles.
\begin{lemma}
The asymptotic growth rate of the random ensemble is given by
\begin{equation}
\lambda_\ell = H(\ell), \quad 0 \le \ell \le 1, 
\end{equation}
where $H(x)$ is the binary entropy function defined by
\begin{equation}
H(x) = -x \log_2 x -(1-x) \log_2 (1-x).
\end{equation}
(Proof)
Substituting the average SS weight distribution (\ref{nonextend}) into expression (\ref{asymptoSSdef}), we have
\begin{eqnarray} \nonumber
\lambda_\ell&=& \lim_{n \rightarrow \infty} \frac 1 n \log_2 S_{\ell n}^{\Ren_{(1-R)n,n}} \\ \nonumber
&=& \lim_{n \rightarrow \infty} \frac 1 n \log_2 {n \choose \ell n} (1 - \ell n 2^{-\ell n})^{(1-R)n} \\ \nonumber
&=& \nonumber 
 \lim_{n \rightarrow \infty} \frac 1 n \log_2 {n \choose \ell n}  \\ \nonumber
&+& \lim_{n \rightarrow \infty} \frac 1 n \log_2(1 - \ell n 2^{-\ell n})^{(1-R)n} \\  \nonumber
&=& H(\ell)  + (1-R) \lim_{n \rightarrow \infty} \log_2(1 - \ell n 2^{-\ell n})\\ 
&=& H(\ell).
\end{eqnarray}
Note that in deriving the fourth equality from the third equality, the following equality
\begin{equation}
\frac 1 n \log_2 {n \choose \ell n} = H(\ell) + o(1),
\end{equation}
was used,  where $o(1)$ denotes terms which converge to $0$ in the limit as $n \rightarrow \infty$.
\hfill\qed
\end{lemma}

\subsubsection{Constant row weight ensemble}
The constant row weight ensemble $\Cen_{m,n,r}$ consists of
all the binary $m \times n$ matrices whose rows have exactly weight $r$ $(m,n \ge 1, r \ge 1)$. 
The size of the ensemble is, thus, 
\[
\# \Cen_{m,n,r} = {n \choose r} ^m.
\]
The average weight distribution of this ensemble was shown in \cite{LS02}.

The following lemma shows the average SS distribution of $\Cen_{m,n,r}$. The
proof is similar to that of Lemma \ref{AveSSrand}.
\begin{lemma}
The average SS distribution of the constant row weight ensemble $\Cen_{m,n,r}$ is given by 
\begin{equation}
S_w^{\Cen_{m,n,r}}  =  {n \choose w}\left( 1- w \frac{{n-w \choose r -1}}{{n \choose r}} \right) ^m, 
\end{equation}
for $0 \le w \le n$. \\
(Proof)
Combining 
\begin{eqnarray}
\nonumber
\# \{H \in \Cen_{m,n,r}:  q_H(\ve x_w) = 0 \} \hspace{3cm} \\
= (\# \{ \ve h \in Z^{(n,r)}: \ve h \circ \ve x_w \ne 1\})^m
\end{eqnarray}
and 
\begin{equation}
\# \{ \ve h \in Z^{(n,r)}: \ve h \circ \ve x_w \ne 1\} = {n \choose r} - w {n-w \choose r -1},
\end{equation}
we have 
\begin{equation}
\# \{H \in \Cen_{m,n,r}:  q_H(\ve x_w) = 0 \} = \left({n \choose r} - w {n-w \choose r -1} \right) ^m.
\end{equation}
The average SS weight distribution is thus given by
\begin{eqnarray} \nonumber
S_w^{\Cen_{m,n,r}} \hspace{-2mm}
&=&\frac{1}{\# \Cen_{m,n,r}}\sum_{\ve x \in Z^{(n,w)}} \hspace{-3mm} \# \{H \in G:  q_H(\ve x) = 0 \} \\ \nonumber
&=& \frac{1}{{n \choose r}^m}{n \choose w}  \left({n \choose r} - w {n-w \choose r -1} \right) ^m \\
&=&  {n \choose w}\left( 1- w \frac{{n-w \choose r -1}}{{n \choose r}} \right) ^m.
\end{eqnarray}
 In the above, the symmetric property of the ensemble is used in deriving the second equality from the first.
\hfill\qed
\end{lemma}

The asymptotic growth rate of the constant row weight ensemble is defined by
\begin{equation} \label{constasymptoSSdef}
\xi_\ell(R, r) \defeq \lim_{n \rightarrow \infty} \frac 1 n \log_2 S_{\ell n}^{\Cen_{(1-R)n,n,r}},
\end{equation}
for  $0 < R < 1$  and $0 \le \ell \le 1$. The next lemma gives the explicit form of the 
average growth rate:
\begin{lemma}
The asymptotic SS weight distribution of the constant row weight ensembles is given by
\begin{equation}
\xi_\ell(R, r)= H(\ell) -(1-R)\log_2\left(\frac{1}{1- r \ell(1-\ell)^{r-1}} \right)
\end{equation}
for  $0 < R < 1$  and $0 \le \ell \le 1$.  \\
(Proof)
By using the lower and upper bounds on binomial coefficients given by
\begin{equation}
\frac{n^k}{k!} \exp\left(-\frac{k^2}{n} \right) \le {n \choose k} \le \frac{n^k}{k!},
\end{equation}
we obtain
\begin{eqnarray} \nonumber
\ell n \frac{{n- \ell n \choose r -1}}{{n \choose r}} &\ge& \frac{ \ell n r (n-\ell n)^{r-1}}{n^r} 
\exp\left(-\frac{(r-1)^2}{n-\ell n} \right) \\ \nonumber 
&=& r \ell (1-\ell)^{r-1}\exp\left(-\frac{(r-1)^2}{n-\ell n} \right) \\
\end{eqnarray}
and
\begin{eqnarray} \nonumber
\ell n \frac{{n-\ell n \choose r -1}}{{n \choose r}} &\le& \frac{\ell n r (n-\ell n)^{r-1}}{n^r} 
\exp\left(\frac{r^2}{n} \right) \\
&=& r \ell (1-\ell)^{r-1}\exp\left(\frac{r^2}{n} \right).
\end{eqnarray}
These bounds imply that
\begin{equation}
 \ell n \frac{{n-\ell n \choose r -1}}{{n \choose r}}
= r \ell (1-\ell)^{r-1}  + o(1)
\end{equation}
since $r$ is constant ({\em i.e.} not a function of $n$). Using this equation, we obtain immediately the
asymptotic SS weight distribution,
\begin{eqnarray} \nonumber
\xi_\ell(R, r) &=& \lim_{n \rightarrow \infty} \frac 1 n \log_2 S_{\ell n}^{\Cen_{(1-R)n,n,r}} \\ \nonumber
&=& \lim_{n \rightarrow \infty} \frac 1 n  
\log_2{n \choose \ell n}\left( 1- \ell n \frac{{n-\ell n \choose r -1}}{{n \choose r}} \right) ^{(1-R)n} \\ \nonumber
&=& H(\ell) + (1-R) \lim_{n \rightarrow \infty}
\log_2  \left(1 -    \ell n \frac{{n-\ell n \choose r -1}}{{n \choose r}}\right) \\
&=& H(\ell) + (1-R) \log_2 (1- r \ell (1-\ell)^{r-1} ).
\end{eqnarray}
\hfill\qed
\end{lemma}

\subsubsection{Bipartite ensemble}
The bipartite graph ensemble (abbreviated as a {\em bipartite ensemble}) $\Ben_{n,c,d}$ is the 
ensemble of regular bipartite graphs of variable node degree $c$ and check node degree $d$.
\footnote{Strictly speaking, to define the average SS weight distribution of  $\Ben_{n,c,d}$, 
we need a graph-based definition of
the stopping sets and ensemble average. Details can be found in \cite{Orlitsky}.}
The following lemma is due to Orlitsky et al\cite{Orlitsky}. 
\begin{lemma}[Orlitsky et al]
The average SS weight distribution of $\Ben_{n,c,d}$ is given by
\begin{equation}
S_w^{\Ben_{n,c,d}} = \frac{\mbox{coef}\left[ ((1+x)^d-dx)^{\frac{c}{d} n}, x^{wc} \right]  }{{nc \choose wc}} {n \choose w},
\end{equation}
where $\mbox{coef}[f(x),x^i]$ denotes the coefficient of a polynomial $f(x)$ corresponding to the term $x^i$.
\hfill\qed
\end{lemma}

The asymptotic growth rate of  $(c,d)$-bipartite ensemble is defined by
\begin{equation}
\gamma_\ell(c,d) \defeq\lim_{n \rightarrow \infty} \frac 1 n \log_2 S_{\ell n}^{\Ben_{n,c,d}}
\end{equation}
for $0 \le \ell \le 1$.
It is shown in \cite{Orlitsky} that the asymptotic growth rate $\gamma_\ell(c,d)$ has the form
\begin{equation}
\gamma_\ell(c,d) = \frac c d \log_e \left(\frac{(1+x_0)^d- d x_0 }{x_0^{\ell d}} \right) -(c-1) H_e(\ell),
\end{equation}
where $x_0$ is the only positive solution of
\begin{equation}
\frac{x ((1+x)^{d-1}- 1)}{(1+x)^d - d x} = \ell
\end{equation}
and $H_e(x)$ is the entropy function with base $e$ defined by
\begin{equation}
H_e(x) \defeq -x \log_e (x) -(1-x) \log_e (1-x).
\end{equation}

\section{Average SS weight distributions of redundant random ensemble}
In this section, we discuss the average SS weight distributions of extended ensembles obtained from 
the random ensemble.

\subsection{Redundant extension}

Before commencing a discussion of redundant extensions, it is perhaps worthwhile to consider 
how some stopping sets can be eliminated by extending a matrix.
\begin{example}
Consider the matrix
\begin{equation}
H \defeq
\left(
\begin{array}{cccc}
0 & 1 & 1 & 1 \\
0 & 1 & 1 & 0 \\
1 & 0 & 1 & 1 \\
\end{array}
\right).
\end{equation}
It is easy to see that $\{2,3,4 \}$ is a stopping set (the sub-matrix composed of the second, third and fourth columns has no row of weight 1). Appending (0 0 0 1) (obtained by adding the first and
second rows of $H$) to $H$ as a row vector, we have a modified matrix $H'$, 
\begin{equation}
H' \defeq
\left(
\begin{array}{cccc}
0 & 1 & 1 & 1 \\
0 & 1 & 1 & 0 \\
1 & 0 & 1 & 1 \\
\hline
0 & 0 & 0 & 1 \\
\end{array}
\right).
\end{equation}
We can observe that 
the weight of the last row of the sub-matrix corresponding to the second, third and fourth columns 
is 1. This implies that $\{2,3,4 \}$ is no longer a stopping set of $H'$. 
Note also that the row spaces spanned by $H$ and $H'$ are
exactly the same.
\hfill\qed
\end{example}

The previous example demonstrates the possibility that
the SS weight distribution could be improved by adding 
linearly dependent rows (called {\em redundant rows}) 
to a given matrix\footnote{It is evident, from the definition of SS,  that the addition of redundant rows
does not introduce a new SS which is a  non-SS of the original matrix.}.

Let $H$ be a  binary $m \times n$ $(1 \le m < n)$ matrix,
\begin{equation}
H \defeq
\left(
\begin{array}{c}
\ve h_1 \\
\ve h_2 \\
\vdots \\
\ve h_m \\
\end{array}
\right).
\end{equation}
Let $L$ be a positive integer which is a divisor of  $m$.
For $1 \le i \le 2^L-1$, $0 \le \ell \le m/L-1$, we define $\ve a_i^{(\ell)}$ by 
\begin{equation}
\ve a_i^{(\ell)} \defeq (b_1(i),b_2(i),\ldots, b_L(i)) 
\left(
\begin{array}{c}
\ve h_{L\ell + 1} \\
\ve h_{L\ell + 2} \\
\vdots \\
\ve h_{L\ell + L} \\
\end{array}
\right),
\end{equation}
where $b_j(i)$ is the $j$-th bit of binary representation of $i$, namely,
$
i = \sum_{j=1}^L 2^j b_j(i).
$
In other words,
$\ve a_i^{(\ell)}$ is a linear combination of $\ve h_{L\ell + k} (1 \le k \le L)$.

The redundant extension of a given matrix is defined as follows:
\begin{definition}[Redundant extension]
The {\em redundant extension} of $H$, denoted by $H^{(L)} $,
 is the matrix whose row vectors are $\ve a_i^{(\ell)}$ 
 for $1 \le i \le 2^L-1$ and $0 \le \ell \le m/L-1$. In other words, $H^{(L)}$ is given by
\begin{equation}
H^{(L)} = 
\left(
\begin{array}{c}
\ve h'_1 \\
\ve h'_2 \\
\vdots \\
\ve h'_{(2^{L}-1)(m/L)} \\
\end{array}
\right),
\end{equation}
where $\ve h'_{(2^L - 1)\ell + i} = \ve a_i^{(\ell)}$ for $1 \le i \le 2^L-1$ and $0 \le \ell \le m/L-1$.
The parameter $L$ is called {\em extension degree}.
The number of row vectors in $H^{(L)}$ is $(2^{L}-1)(m/L)$.
\hfill\qed
\end{definition}
\begin{figure}[htbp]
  \begin{center}
  \includegraphics[scale=0.45]{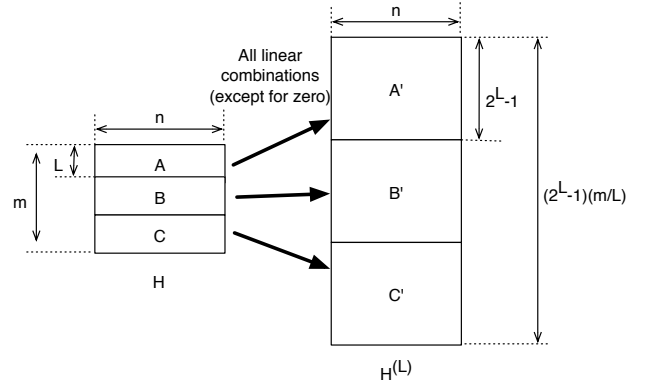} \\
\begin{flushleft}
{\small  A sub-block of $H$ 
corresponds to a sub-block in $H^{(L)}$ (for example, sub-block A corresponds to sub-block A').  
The row vectors in sub-block X' (X $\in$ \{A,B,C\} ) can be obtained by constructing all linear combinations
(except for the zero combination) of the row vectors in sub-block X.
}
\end{flushleft}
      \caption{Redundant extension of a parity check matrix. }
    \label{fig:extension}
      \end{center}
\end{figure}
Figure \ref{fig:extension} illustrates the idea of the redundant extension.

From the above definition of the redundant extension, it is clear that 
the row spaces  of $H$ and $H^{(L)}$  are the same. In other words,
the code defined by $H$ coincides with  the code defined by $H^{(L)}$.
However, although the codes defined by $H$ and  $H^{(L)}$ are the same, 
$H$ and $H^{(L)}$ may have different  SS weight distributions.

\begin{example}
For a given $m \times n$ matrix $H$,  $H^{(2)} $ is expressed as 
\begin{equation}
H^{(2)} = 
\left(
\begin{array}{c}
\ve h_1 \\
\ve h_2 \\
\ve h_1 \oplus \ve h_2 \\
\ve h_3 \\
\ve h_4 \\
\ve h_3 \oplus \ve h_4 \\
\vdots \\
\ve h_{m-1}\\
\ve h_m \\
\ve h_{m-1} \oplus \ve h_m \\
\end{array}
\right).
\end{equation}
\hfill\qed
\end{example}

The definition of the redundant extension of a matrix naturally leads to the following definition of
the redundant extension of a given ensemble: 
\begin{definition}[Extended ensemble]
Consider the case where a random ensemble  $\Gen$ which consists of $m \times n$ binary matrices is given.
Let $L$ be a divisor of $m$. 
The extended ensemble of $\Gen$, denoted by $\Gen^{(L)}$, is defined by 
\begin{equation}
\Gen^{(L)} \defeq \{ H^{(L)}: H \in \Gen\}.
\end{equation}
The size of the ensemble $\# \Gen^{(L)}$ is equal to the size of the original ensemble $\# \Gen$.
An equal probability  is assigned
to each matrix in $\Gen^{(L)}$.
\hfill\qed
\end{definition}
The {\em redundant random ensemble} which is the main subject of this paper is the
extended ensemble of a random ensemble, which is
denoted by  $\Ren_{m,n}^{(L)}$.

\subsection{Redundant random ensemble: $L = 2$}

In this subsection, we discuss the average SS weight distribution of
the redundant random ensemble  $\Ren_{m,n}^{(2)}$.
In this case, we can derive a simple exact formula for the average SS weight distribution.

Suppose that the first 3-rows of $H \in \Ren_{m,n}^{(2)}$, 
\begin{equation}
\tilde H \defeq
\left(
\begin{array}{c}
\ve h_1 \\
\ve h_2 \\
\ve h_1 \oplus \ve h_2 \\
\end{array}
\right),
\end{equation}
are given. Our first task is to count the number of 
pairs $(\ve h_1, \ve h_2)\in F_2^n \times F_2^n$ satisfying  $q_{\tilde H}(\ve x_w) = 0$.
Let us define $U$ by
\begin{eqnarray} \nonumber
U \defeq \# \{(\ve h_1,\ve h_2) \in F_2^n \times F_2^n: \ve h_1 \circ \ve x_w \ne 1, \ve h_2 \circ \ve x_w \ne 1, \\
\label{Udef}
(\ve h_1 \oplus \ve h_2) \circ \ve x \ne 1   \} .
\end{eqnarray}
Note that $m/2$ sub-blocks exist in $H \in \Ren_{n,m}^{(2)}$, and
these sub-blocks can be chosen independently when we count 
$\#\{H \in \Ren_{n,m}^{(2)}: q_H(\ve x_w) = 0 \}$. This observation leads to the following equality,
\begin{equation} \label{4theq}
\#\{H \in \Ren_{n,m}^{(2)}: q_H(\ve x_w) = 0 \} = U^{m/2}.
\end{equation}
The next lemma gives a simple description of $U$.
\begin{lemma} 
\label{Ulemma}
For $1 \le m < n$,  $ w \ge 1$,  $U$ is given by
\begin{equation}
U = (2^n - w 2^{n-w})^2  - 2^{2(n-w)+1} \sum_{\gamma=2}^w {w \choose \gamma} (w - \gamma) .
\end{equation}
(Proof)
Let 
\begin{eqnarray} \nonumber
V \defeq \# \{(\ve h_1,\ve h_2) \in F_2^n \times F_2^n: \ve h_1 \circ \ve x_w \ne 1, \ve h_2 \circ \ve x_w \ne 1, \\
(\ve h_1 \oplus \ve h_2) \circ \ve x_w = 1   \}.
\end{eqnarray}
Using the equality
\begin{eqnarray} \nonumber
\# \{(\ve h_1, \ve h_2) \in F_2^n \times F_2^n: \ve h_1 \circ \ve x_w \ne 1,\ve h_2 \circ \ve x_w \ne 1 \} \\
=(2^n - w 2^{n-w})^2,
\end{eqnarray}
we have
\begin{eqnarray} \nonumber
U &=& \{(\ve h_1,\ve h_2) \in F_2^n \times F_2^n:
 \ve h_1 \circ \ve x_w \ne 1,\ve h_2 \circ \ve x_w \ne 1 \} \\ \nonumber
 &-& V \\ \label{Ueq}
&=& (2^n - w 2^{n-w})^2 - V.
\end{eqnarray}
In the following, we will evaluate $V$.
Define $\alpha,\beta, \gamma$ by
\begin{eqnarray}
\alpha &\defeq& \# \{ i \in [1,n]: h_{i}^{(1)} = 1, h_{i}^{(2)} = 0, x_i = 1\} \\
\beta &\defeq& \# \{ i \in [1,n]: h_{i}^{(1)} = 0, h_{i}^{(2)} = 1, x_i = 1\} \\
\gamma &\defeq& \# \{ i \in [1,n]: h_{i}^{(1)} = 1, h_{i}^{(2)} = 1, x_i = 1\}, 
\end{eqnarray}
where $\ve h_1 = (h_1^{(1)},h_2^{(1)},\ldots, h_n^{(1)} )$ and 
$\ve h_2 = (h_1^{(2)},h_2^{(2)},\ldots, h_n^{(2)} )$.
Assume that $\ve h_1 \circ \ve x_w \ne 1$ and $\ve h_2 \circ \ve x_w \ne 1$.
In this case, the equality $(\ve h_1 \oplus \ve h_2) \circ \ve x_w = 1$
holds if and only if the following two  conditions hold:
(i) $(\alpha,\beta)=(1,0)$ or $(\alpha,\beta)=(0,1)$,
(ii) $2 \le \gamma \le w$.
Suppose the case $(\alpha,\beta)=(1,0)$.
The number of possible pairs $(\ve h_1, \ve h_2)$ satisfying
$\ve h_1 \circ \ve x_w \ne 1, \ve h_2 \circ \ve x_w \ne 1$,
$(\ve h_1 \oplus \ve h_2) \circ \ve x_w = 1$
is given by
\begin{equation}
\sum_{\gamma=2}^w {w \choose \gamma} (w - \gamma) 2^{2(n-w)}.
\end{equation}
Taking the case $(\alpha, \beta) = (0,1)$ into consideration, we immediately have
\begin{equation}
V = 2^{2(n-w)+1} \sum_{\gamma=2}^w {w \choose \gamma} (w - \gamma).
\end{equation}
Substituting this equation into Eq.(\ref{Ueq}), we obtain the claim of the lemma.
\hfill\qed
\end{lemma}

The following theorem is an immediate consequence of Lemma \ref{Ulemma}. 
\begin{theorem}[Average SS distribution of $\Ren_{m,n}^{(2)}$ ]
\label{thave2}
The average SS weight distribution of $\Ren_{m,n}^{(2)}$  is given by
\begin{eqnarray}
S_w^{\Ren_{m,n}^{(2)}} &=& \frac{{n \choose w}}{2^{mn}}\left( (2^n - w 2^{n-w})^2  -V  \right)^{m/2} 
\end{eqnarray}
for $1 \le m < n$,  $ w \ge 1$, where $V$ is defined by
\begin{equation}
V \defeq  2^{2(n-w)+1} \sum_{\gamma=2}^w {w \choose \gamma} (w - \gamma).
\end{equation}
 \\
(Proof)
The average SS weight distribution of $\Ren_{n,m}^{(2)}$ can be derived in the following way:
\begin{eqnarray} \nonumber
S_w^{\Ren_{m,n}^{(2)}}  \hspace{-2mm}
&=& \frac{1}{\# \Ren_{m,n}^{(2)}}\sum_{\ve x \in Z^{(n,w)}} \hspace{-4mm}
  \#\{H \in \Ren_{n,m}^{(2)}: q_H(\ve x) = 0 \} \\ \nonumber
&=& \frac{{n \choose w}}{2^{mn}}    \#\{H \in \Ren_{m,n}^{(2)}: q_H(\ve x_w) = 0 \} \\ \nonumber
&=& \frac{{n \choose w}}{2^{mn}}   U^{m/2} \\
&=& \frac{{n \choose w}}{2^{mn}}\left( (2^n - w 2^{n-w})^2  - V  \right)^{m/2}
\end{eqnarray}
The second equality follows from the symmetric property of the ensemble, while
the third equality is derived from Eq.(\ref{4theq}). The last equality is due to Lemma \ref{Ulemma}.
\hfill\qed
\end{theorem}

\begin{example}
Consider the case where $n = 4, m = 2$. The ensemble $\Ren_{2,4}^{(2)}$ consists of 
 matrices of the form 
\begin{equation}
\left(
\begin{array}{c}
\ve h_1 \\
\ve h_2 \\
\ve h_1 \oplus \ve h_2 \\
\ve h_3 \\
\ve h_4 \\
\ve h_3 \oplus \ve h_4 \\
\end{array}
\right).
\end{equation}
We consider the case that
the row vectors $\ve h_1, \ve h_2, \ve h_3, \ve h_4$ are chosen from $F_2^4$ 
with uniform probability.
From Theorem \ref{thave2}, we have
\begin{equation}
\left\{S_w^{\Ren_{2,4}}\right\}_{w=1}^4
 = \left\{1,\frac 3 2, \frac{19}{16}, \frac{7}{16} \right\} .
 \end{equation}
On the other hand, $\Ren_{2,4}$, which is the set of matrices of the form
\begin{equation}
\left(
\begin{array}{c}
\ve h_1 \\
\ve h_2 \\
\ve h_3 \\
\ve h_4 \\
\end{array}
\right),
\end{equation}
has the average SS weight distribution
\begin{equation}
\left\{S_w^{\Ren_{2,4}}\right\}_{w=1}^4
 = \left\{1,\frac 3 2, \frac{25}{16}, \frac{9}{16} \right\} ,
 \end{equation}
where this distribution is derived using Lemma \ref{AveSSrand}. It can be observed that
the average SS weights of the extended ensemble are smaller those that of the original 
ensemble in the cases $w = 3$ and $4$. \hfill\qed
\end{example}
The argument used in the proof of Theorem \ref{thave2} can be used to derive 
the average SS weight distribution of the extended constant row weight ensemble 
with $L = 2$, the details are summarized in the Appendix. Here we consider the following example.
\begin{example}
Let $n = 100, m = 50$.
The average SS weight distribution of the extended constant row weight
ensembles with $L = 2$, $S_w^{\Cen_{100,50,r}^{(2)}} $, can be evaluated  using
Lemma \ref{lemmapp} of the Appendix. Tables \ref{tb1} and \ref{tb2} present
the two cases $r = 10$ (sparse matrix) and $r=50$ (dense matrix),  respectively.
We can see that the improvement due to extension  is very small 
for the case $r = 10$. On the other hand, a significant improvement can be observed for the
case $r=50$.
\hfill\qed
\begin{table}[htdp]
\caption{Comparison of a non-extended ensemble and an extended ensemble(constant row 
weight ensemble, sparse case).}
\label{tb1}
\begin{center}
\begin{tabular}{rrr}
\hline
\hline
$w$ & $S_w^{\Cen_{n,m,r}} $ & $S_w^{\Cen_{n,m,r}^{(2)}} $ \\
\hline
1 & 0.515 & 0.515 \\
2 & 0.217 & 0.217 \\
3 & 0.107 & 0.107 \\
4 & 0.0726 & 0.0721 \\
5 & 0.0748 & 0.0737 \\
6 & 0.123 & 0.119 \\
7 & 0.322 & 0.308 \\
8 & 1.33 & 1.24 \\
9 & 8.20 & 7.54 \\
10 & 71.5 & 64.6 \\
\hline
\end{tabular} \\
$n = 100, m = 50, r = 10$
\end{center}
\label{lowdensity}
\end{table}%
\begin{table}[htdp]
\caption{Comparison of  a non-extended ensemble  and an extended ensemble(constant row 
weight ensemble, dense case).}
\label{tb2}
\begin{center}
\begin{tabular}{lll}
\hline
\hline
$w$ & $S_w^{\Cen_{n,m,r}} $ & $S_w^{\Cen_{n,m,r}^{(2)}} $ \\
\hline
1 & $8.88 \times 10^{-14}$ & $8.88 \times 10^{-14}$ \\
2 & $2.65 \times 10^{-12}$ & $2.65 \times 10^{-12}$ \\
3 & $7.43 \times 10^{-6}$ & $8.32 \times 10^{-9}$ \\
4 & $2.23\times 10^0$  & $4.18 \times 10^{-3}$ \\
5 & $1.87 \times 10^4$ & $3.08 \times 10^2$ \\
\hline
\end{tabular} \\
$n = 100, m = 50, r = 50$
\end{center}
\label{highdensity}
\end{table}%
\end{example}
This example suggests that
the advantages of redundant extension are more significant when
the original matrix is dense.
This is one of the major reasons that we focus on 
redundant random ensembles in the present study.

\subsection{Redundant random ensemble: $L > 2$}

It becomes difficult to evaluate the number of extended parity check matrices giving
 a stopping set for a given weight when $L > 2$.Instead of deriving an exact expression,
we here utilize upper and lower bounds on the number of such parity check matrices to
study the average SS weight distribution of extended ensembles.

\subsubsection{Number of generator matrices with minimum distance greater than or equal to 2}

Let $G$ be a binary $K \times N$ matrix($1\le K < N$).
The weight distribution of the code generated from $G$ (as a generator matrix) is 
defined by
\begin{equation}
A_w(G) \defeq \sum_{\ve c \in Z^{(N,w)}}\sum_{\ve m \in \bar F_2^{K}}I[\ve m G= \ve c]
\end{equation}
for $1 \le w \le N$, where $\bar F_2^{K}$ denotes $F_2^K \backslash \ve 0$.
The average of the weight distribution (where expectation is taken over a given ensemble $\Gen$) is given by
\begin{eqnarray}
\hspace{-3mm}
E_\Gen[A_w(G)] \hspace{-2mm} &=& \hspace{-3mm}\sum_{G \in \Gen} P(G) \sum_{\ve c \in Z^{(N,w)}}
\sum_{\ve m \in \bar F_2^K}I[\ve m G= \ve c] \\
&=&  \hspace{-3mm}\frac{1}{\# \Gen}\hspace{-2mm} \sum_{\ve c \in Z^{(N,w)}}\sum_{\ve m \in \bar F_2^K}  
\hspace{-3mm} \#\{ G \in \Gen: \ve m G= \ve c\}.
\end{eqnarray}
The minimum distance of $G$ is given by
\begin{equation}
d_{min}(G) \defeq \min \{ i \in [1,n]: A_i(G) \ne 0 \}.
\end{equation}

In order to prove the upper and lower bounds on the number of certain parity check matrices, we can use
the first and second moment method\cite{alon}, which requires the first and second moments of a 
random variable.
The following lemma is presented in the problem section of \cite{modern} (the proof is given in the
Appendix).
\begin{lemma}
\label{1st2nd}
The first and second moments of $A_w(G)$ with respect to $\Ren_{K,N}$ are given by
\begin{equation} \label{1strandom}
E_{\Ren_{K,N}}[A_w(G)] = (2^K - 1) 2^{-N} {N \choose w}
\end{equation}
and
\begin{eqnarray} \nonumber
E_{\Ren_{K,N}}[A_w(G)^2] &=& E_{\Ren_{K,N}}[A_w(G)] ^2 +E_{\Ren_{K,N}}[A_w(G)] \\ \label{2ndrandom}
&\times &  \left(1 - {N \choose w} 2^{-N} \right)
\end{eqnarray}
respectively, for $w  \in [1,n]$.
\hfill\qed
\end{lemma}

The next lemma is the basis of the lower bound on the average SS weight distribution for redundant random ensembles.
\begin{lemma}
\label{lowernum}
The number of matrices in $\Ren_{K,N}$ which have minimum distance greater than or equal to 2 has a lower bound given by

\[
\# \{G \in \Ren_{K,N}: d_{min}(G) \ge 2  \} \hspace{4cm}
\]
\begin{equation}
\ge
\max\{2^{KN}\max\{1 -(2^K - 1)2^{-N} N,0\}, 2^{KN-K} \} .
\end{equation}
(Proof) 
Let $D \defeq \# \{G \in \Ren_{K,N}: d_{min}(G) \ge 2  \}$.
We first prove $D \ge 2^{KN}\max\{1 -(2^K - 1)2^{-N} N,0\}$.
The number of matrices whose minimum distance is greater than or equal to 2 
can be written in the form
\begin{eqnarray} \nonumber
D 
&=& \# \{G \in \Ren_{K,N}: A_1(G) = 0  \} \\ \nonumber
&=& \# \{G \in \Ren_{K,N}: A_1(G) < 1  \} \\ \nonumber
&=& \sum_{G \in \Ren_{K,N} } I[A_1(G) < 1] \\ \nonumber
&=& 2^{KN}\sum_{G \in \Ren_{K,N} } P(G)I[A_1(G) < 1] \\ \label{a1g}
&=& 2^{KN} Pr[A_1(G) < 1],
\end{eqnarray}
where $Pr[A_1(G) < 1]$ is given by 
\begin{equation}
Pr[A_1(G) < 1] = \sum_{G \in \Ren_{K,N} } P(G)I[A_1(G) < 1] .
\end{equation}
The Markov inequality implies 
\begin{equation}
Pr[A_1(G) \ge 1] \le E_{\Ren_{K,N}}[A_1(G)], 
\end{equation}
which is equivalent to
\begin{equation}
Pr[A_1(G) < 1] \ge \max\{1 - E_{\Ren_{K,N}}[A_1(G)] ,0\}.
\end{equation}
Substituting this upper bound into Eq.(\ref{a1g}) and using Eq.(\ref{1strandom}), we have
\begin{eqnarray}\nonumber
D &\ge&  2^{KN}\max\{1 - E_{\Ren_{K,N}}[A_1(G)] ,0\}\\
&=&2^{KN} \max\{1 -(2^K - 1)2^{-N} N,0 \}.
\end{eqnarray}
We then consider the inequality $D \ge 2^{KN-K}$.  Suppose the case
that every row of $G' \in \Ren_{K,N}$ is of even weight. We call this condition 
the even weight condition. In such a case, no linear
combination of rows of $G'$ gives a vector of weight 1. The number of $K \times N$ 
matrices satisfying the even weight condition is $2^{K(N - 1)}$ because
there exist $2^{N-1}$ even weight vectors of length $N$. 
\hfill\qed
\end{lemma}

The next lemma will be required to prove an upper bound on the average SS weight distribution 
of redundant random ensembles.
\begin{lemma}\label{numupper}
The number of matrices in $\Ren_{K,N}$ which have minimum distance greater than or equal to 2 has an upper bound given by

\[
\# \{G \in \Ren_{K,N}: d_{min}(G) \ge 2  \} \hspace{4cm}
\]
\begin{equation}
\le 2^{KN} \frac{1-N2^{-N}}{(2^K - 1) N2^{-N} + 1-N2^{-N}}.
\end{equation}
(Proof)
For a non-negative integer-valued random variable $X$,
the following inequality holds\cite{alon},
\[
Pr[X = 0] \le \frac{E[X^2]-E[X]^2}{E[X^2]}.
\]
Considering $A_1(G)$ as a random variable, we obtain 
\[
\# \{G \in \Ren_{K,N}: d_{min}(G) \ge 2  \}  \hspace{60mm} 
\]
\vspace{-8mm}
\begin{eqnarray} \nonumber
&=&\# \{G \in \Ren_{K,N}:  A_1(G) = 0  \}  \\ \nonumber
&=& 2^{KN} Pr[A_1(G) = 0]  \\ \label{d2upper}
&\le& 2^{KN} \frac{E_{\Ren_{K,N}}[A_1(G)^2]-E_{\Ren_{K,N}}[A_1(G)]^2}{E_{\Ren_{K,N}}[A_1(G)^2]}.
\end{eqnarray}
From Lemma \ref{1st2nd}, the first and the second moments of $A_1(G)$ are given by
\begin{eqnarray}
E_{\Ren_{K,N}}[A_1(G)] \hspace{-2mm} &=& \hspace{-2mm}(2^K - 1) N2^{-N},\\  \nonumber
E_{\Ren_{K,N}}[A_1(G)^2]  &=& \hspace{-2mm} ((2^K - 1) N2^{-N})^2  \\
&+&  \hspace{-3mm}(2^K - 1) N2^{-N} \left(1 - N 2^{-N} \right).
\end{eqnarray}
Substituting these expressions into Eq.(\ref{d2upper}), we have the claim of the lemma.
\hfill\qed
\end{lemma}

\subsubsection{Upper and lower bounds on average SS weight distributions}
We are now ready to prove the following upper and lower bounds on the average SS weight distribution.
\begin{theorem}[Upper and lower bounds on $S_w^{\Ren_{m,n}^{(L)}}$ ]
\label{upperlowerL}
The average SS weight distribution of the redundant random 
ensemble satisfies the inequalities
\begin{eqnarray}\label{lowerrandom}
S_w^{\Ren_{m,n}^{(L)}}  
&\ge & 
 {n \choose w} \max\{A^{m/L},  2^{-m}\}   \\
S_w^{\Ren_{m,n}^{(L)}}  
&\le&  {n \choose w}\left(\frac{1-w2^{-w}}{(2^L - 1) w2^{-w}+1-w2^{-w}}    \right)^{m/L} 
\end{eqnarray}
for $1 \le w \le n$, where $A$ is defined by
\begin{equation}
A \defeq \max\{1 -(2^L - 1)2^{-w} w,0 \}.
\end{equation}
(Proof)
From the definition of the average SS weight distribution, 
$S_w^{\Ren_{m,n}^{(L)} }$ can be expressed as
\begin{eqnarray} \nonumber
S_w^{\Ren_{m,n}^{(L)} }
\hspace{-3mm}
 &=&  \frac{1}{2^{mn}} \sum_{\ve x \in Z^{(n,w)}}\#  \{H \in \Ren_{m,n}^{(L)} : q_H(\ve x) = 0 \} \\ \nonumber
 &=&  \frac{1}{2^{mn}} {n \choose w}\#  \{H \in \Ren_{m,n}^{(L)} : q_H(\ve x) = 0 \}  \\ \nonumber
&=&  \frac{1}{2^{mn}}  {n \choose w}\left( \# \{H' \in \Ren_{L,n}^{(L)}: q_H(\ve x) = 0 \}  \right)^{m/L}\\ \nonumber
&=&  \frac{1}{2^{mn}}  {n \choose w} \\ \label{lasteq}
&\times& \hspace{-4mm} \left( \# \{G \in \Ren_{L,w}: d_{min}(G) \ge 2\}2^{L(n-w)}  \right)^{m/L},
\end{eqnarray}
where, in the above, the second equality was obtained by using the symmetric property of the ensemble, and the third equality arises from the property 
 that  $L$ sub-blocks can be chosen independently.
The last equality holds because
\[
\# \{H' \in \Ren_{L,n}^{(L)}: q_{H'}(\ve x) = 0 \} \hspace{4cm}
\]
\begin{equation}
= \# \{G \in \Ren_{L,w}: d_{min}(G) \ge 2\}  2^{L(n-w)}.
\end{equation}

Note that $d_{min}(G) \ge 2$ means that 
no linear combination of row vectors of $G$ (except for all zero coefficients) has weight 1.
Applying the inequality in Lemma \ref{numupper} to Eq.(\ref{lasteq}), we immediately obtain the upper bound,
\begin{eqnarray} \nonumber
S_w^{\Ren_{m,n}^{(L)} } 
&=&  \frac{1}{2^{mn}}  {n \choose w} \\ \nonumber
&\times& \left( \# \{G \in \Ren_{L,w}: d_{min}(G) \ge 2\}2^{L(n-w)}  \right)^{m/L} \\ \nonumber
&\le &  \frac{1}{2^{mn}} {n \choose w}\left( 2^{Lw}
\times B  \times 2^{L(n-w)}\right)^{m/L}\\ \nonumber
&= &  \frac{1}{2^{mn}} {n \choose w}\left( B \times  2^{Ln}\right)^{m/L}\\ \label{dminexpr}
&= & {n \choose w} B^{m/L} ,
\end{eqnarray}
where $B$ is defined by
\begin{equation}
B =  \frac{1-w2^{-w}}{(2^L - 1)w 2^{-w}+1-w2^{-w} }.
\end{equation}

On the other hand, applying the inequality in Lemma \ref{lowernum}
to Eq.(\ref{dminexpr}), the lower bound can be derived.
Let $D' \defeq \# \{G \in \Ren_{K,N}: d_{min}(G) \ge 2  \}$.
The lower bound in Lemma \ref{lowernum} is equivalent to
\begin{equation}
D' \ge  \max\{2^{Lw}A ,2^{L w - L} \},
\end{equation}
which leads to the lower bound 
\begin{eqnarray} \nonumber
S_w^{\Ren_{m,n}^{(L)} } 
&=&  \frac{1}{2^{mn}}  {n \choose w}\left( D' 2^{L(n-w)}  \right)^{m/L} \\ \nonumber
&\ge& \frac{1}{2^{mn}} {n \choose w}  
\left( \max\{2^{Lw}A ,2^{L w - L} \} 2^{L(n-w)} \right)^{m/L} \\
&=&
 {n \choose w}\max\{A^{m/L},2^{-m}\}.
\end{eqnarray}
\hfill\qed
\end{theorem}
It is easy to check that 
the upper bound and the lower bound coincide with the average 
SS weight distribution of the non-extended ensemble $S_w^{\Ren_{m,n}} $ given in Eq. (\ref{nonextend})
if $L = 1$.

\begin{example}
Consider the case $n = 100, m = 50, L=2$.
In this case, we can compute exact values of the average SS weight 
distribution due to Theorem \ref{thave2}.
Table \ref{L2table} presents the exact values (Theorem \ref{thave2}) 
together with the values of the upper and lower bound(Theorem \ref{upperlowerL})
of the average SS weight distribution of $\Ren_{50,100}^{(2)}$.
For  the cases $w = 1$ and $2$, we can see that the values of the upper and lower bounds
coincide and they give the exact values.
\begin{table}[htdp]
\caption{Average SS weight distribution $S_w^{\Ren_{50,100}^{(2)} }$: exact, upper and lower bounds.}
\begin{center}
\begin{tabular}{llll}
\hline
\hline
$w$ & Exact & Upper  & Lower \\
\hline
1 & $8.88 \times 10^{-14}$ & $8.88 \times 10^{-14}$ & $8.88 \times 10^{-14}$ \\
2 & $4.40 \times 10^{-12}$ & $4.40 \times 10^{-12}$ & $4.40 \times 10^{-12}$ \\
3 & $1.05 \times 10^{-8}$ & $1.07 \times 10^{-6}$ & $1.44 \times 10^{-10}$ \\
4 & $4.15 \times 10^{-3}$ & $1.17 \times 10^{-1}$ & $3.48 \times 10^{-9}$ \\
5 & $2.58 \times 10^2$ & $1.20 \times 10^3$ & $1.02 \times 10^1$ \\
\hline
\end{tabular}
\end{center}
\label{L2table}
\end{table}%
\hfill\qed
\end{example}

An exact ( non-trivial)  expression for $S_w^{\Ren_{m,n}^{(L)} } (L >2)$ does not at present exist.
Let $Q_{L,w} \defeq \# \{G \in \Ren_{L,w}: d_{min}(G) \ge 2\}$. The source of 
the difficulty in deriving an exact expression comes from the difficulty in counting
$Q_{L,w} $ precisely. However, if both $L$ and $w$ are small, we can 
obtain $Q_{L,w}$ by an exhaustive computer search.
Table \ref{Qlw} presents the values of $Q_{L,w}$ for $1 \le w, L \le 5$ which 
have been evaluated by such an exhaustive computer search.
\begin{table}[htdp]
\begin{center}
\caption{Values of $Q_{L,w}$(The number of matrices with minimum distance greater than or equal to 2).}
\label{Qlw}
\begin{tabular}{c|ccccc}
\hline
\hline
$L  \backslash w$ & 1 & 2 & 3 & 4 & 5\\
\hline
1 & 1 &  2 &  5 & 12 & 27 \\
2 & 1 & 4 & 19 & 112 & 619 \\
3 & 1 & 8 & 71 & 792 & 10683\\
4 & 1 & 16 & 271 & 5416 & 140251\\
5 & 1 & 32 & 1055 & 38472 & 1751067 \\
\hline
\end{tabular}
\end{center}
\end{table}%

The following lemma gives exact value of
$S_w^{\Ren_{m,n}^{(L)} } $ if we know the value of $Q_{L,w}$.
\begin{lemma}
\label{exactexpre}
The  average SS weight distribution of $\Ren_{m,n}^{(L)}$ is given by
\begin{equation}
S_w^{\Ren_{m,n}^{(L)}}   = 
\frac{1}{2^{mn}}  {n \choose w}\left( Q_{L,w} 2^{L(n-w)}  \right)^{m/L}.
\end{equation}
(Proof) The claim of the lemma has already been derived as Eq.(\ref{lasteq}).
\hfill\qed
\end{lemma}

\begin{example}
Consider the case $n = 100, m = 50, L=5$.
Combining Lemma \ref{exactexpre} and the result presented in Table \ref{Qlw},
we can derive the exact values of the average SS weight distribution
for $1 \le w, L \le 5$. These values are presented in Table \ref{tblL5} together with
the corresponding values of the upper and lower bounds.
\begin{table}[htdp]
\caption{Average SS weight distribution $S_w^{\Ren_{50,100}^{(5)} }$: exact, upper and lower bounds.}
\begin{center}
\label{tblL5}
\begin{tabular}{llll}
\hline
\hline
$w$ & Exact & Upper  & Lower \\
\hline
1 & $8.88 \times 10^{-14}$ & $8.88 \times 10^{-14}$ & $8.88 \times 10^{-14}$ \\
2 & $4.40\times 10^{-12}$ & $4.40 \times 10^{-12}$ & $4.40 \times 10^{-12}$ \\
3 & $1.94 \times 10^{-10}$ & $1.93 \times 10^{-8}$ & $1.44 \times 10^{-10}$ \\
4 & $1.73 \times 10^{-8}$ & $1.12 \times 10^{-4}$ & $3.48 \times 10^{-9}$ \\
5 & $1.13  \times 10^{-5}$ & $3.89 \times 10^{-1}$ & $6.69 \times 10^{-8}$ \\
\hline
\end{tabular}
\end{center}
\end{table}%
\hfill\qed
\end{example}

There is a trade-off between the extension degree $L$ and the average SS weight distribution.
 The decoding complexity of BP-based iterative decoding increases as $L$ increases
because the number of rows in the extended matrix $(2^L-1)(m/L)$ is an exponentially increasing function of $L$.
On the other hand,
a large $L$ tends  to give a larger stopping distance. The next example demonstrates such a trade off relation.
\begin{example}
Figure \ref{fig:ul} presents the relation between $L$ and the upper bound of
the average SS weight distribution.  The horizontal axis of Fig.\ref{fig:ul} represents
the weight $w$. The ensemble assumed here is the random ensemble with
$n = 100, m=50$, namely $\Ren_{50,100}$. We can observe that 
the upper bound on the average SS weight distribution 
decreases as  $L$ increases for a fixed weight.
\begin{figure}[htbp]
  \begin{center}
  \includegraphics[scale=0.6]{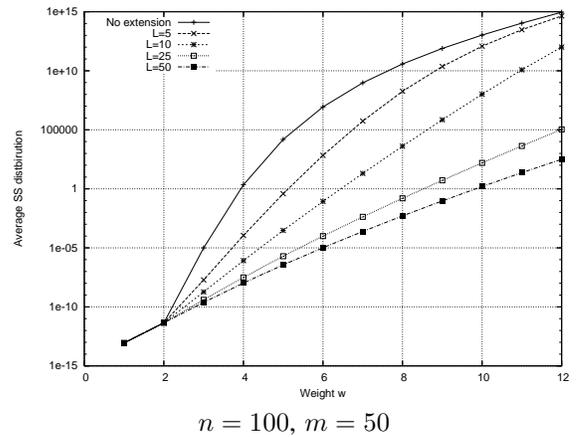} \\
  $n = 100$, $m = 50$ \\
      \caption{Upper  bounds on the average SS weight distribution of redundant random ensembles.}
    \label{fig:ul}
      \end{center}
\end{figure}
\hfill\qed
\end{example}

\begin{example}
Figure \ref{fig:Lcomp} shows the block error probability of three example ensembles
with $m = 50, n=100$:
the random ensemble (matrix A) ,  the redundant random 
ensemble with  $L=2$ (matrix B) and the redundant random ensemble  with
$L=5$(matrix C). The channel is BEC and BP is used in the decoder.
It is observed that the decoding performance of
matrix C is the best among the three matrices. 
The reason for these differing performances can be seen with reference to Table \ref{distmult}. This table presents the stopping distance of
the three matrices and their multiplicity. The multiplicity is the number of the stopping sets
with size equal to the stopping distance. These values have been computed by an
exhaustive computer search.
The matrix C has the largest stopping distance, 7,
which gives a smaller block error probability compared with those of  
matrices A(stopping distance 4) and B(stopping distance 5).
\begin{figure}[htbp]
  \begin{center}
  \includegraphics[scale=0.6]{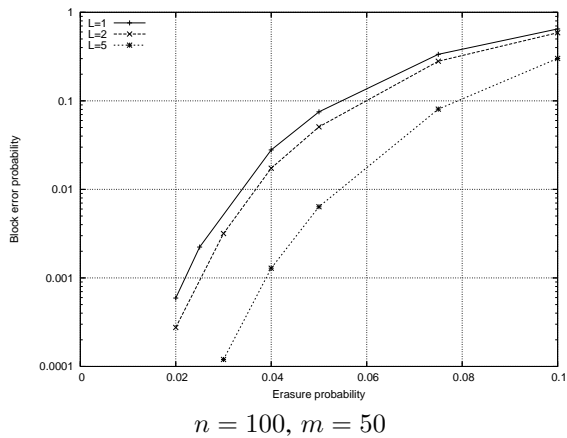} \\
  $n = 100$, $m = 50$ \\
      \caption{Block error probabilities of examples of random 
      and redundant random ensembles (Matrices A,B, and C from top to bottom).}
    \label{fig:Lcomp}
      \end{center}
\end{figure}
\begin{table}[htdp]
\caption{Stopping distance and multiplicity of matrices A,B and C.}
\label{distmult}
\begin{center}
$n=100, m=50$ \\
\begin{tabular}{rrrrr}
\hline
\hline
Matrix & $L$ & Stopping distance & Multiplicity & \# of rows\\
\hline
A &1&  4 & 1 & 50  \\
B &2&  5 & 262 & 75 \\
C &5& 7 & 1365 & 310 \\
\hline
\end{tabular}
\end{center}
\label{matricesABC}
\end{table}%
\hfill\qed
\end{example}

\subsection{Typical stopping distance}

From the average SS weight distribution, we can retrieve some information about
the stopping distance of matrices contained in an ensemble.
\begin{definition}[Typical stopping distance]
The typical stopping distance of an ensemble $\Gen$ is defined by
\begin{equation}
\delta^{\Gen} \defeq \min \left\{ s \in [1,n]: \sum_{w=1}^{s-1} S_w^{\Gen} \ge 1 \right\}.
\end{equation}
\hfill\qed
\end{definition}
The condition $\sum_{w=1}^{s-1}S_w(H) = 0$  
is equivalent to $\Delta(H) \ge s$. It is evident that
there exists a matrix $H \in \Gen$ satisfying $\sum_{w=1}^{\delta^{\Gen}-1}S_w(H) = 0$,
because the average $\sum_{w=1}^{\delta^{\Gen}-1} S_w^{\Gen} $ is strictly smaller than 1.
This means that there exists a matrix with a stopping distance larger than or equal to 
the typical stopping distance $\delta^{\Gen}$.

We here compare a high rate redundant random ensemble with constant row weight ensembles
and bipartite ensembles in terms of their typical stopping distances.
\begin{example}
Consider the case $n = 1024, m = 32$.
We can show that
the maximum value of the typical stopping distance of $\Cen_{n,m,r}$ is
\[
\max_{r \in [1,1024]} \delta^{\Cen_{1024,32,r}} = 3.
\]
For bipartite ensembles, we have
\[
\max_{c\ge 3} \delta^{\Ben_{c,32 c}} = 3.
\]
These results mean that there are no constant row weight ensembles and bipartite ensembles
with $n = 1024, m=32$ which achieve the typical stopping distance 4.
On the other hand, the redundant random ensemble $ (n = 1024,m=32, L=8)$ has a
larger typical stopping distance:
\[
 \delta^{\Ren_{32,1024}^{(8)}} = 4.
\]
In this case, the extended random ensemble is expected to give asymptotically ({\em i.e.}, in the regime of small erasure probability)
better decoding performance (with BP) than the constant row weight ensemble with any row weight and
the bipartite ensemble.

Figure \ref{fig:constrandom} presents the block error probabilities of examples of
a redundant random ensemble ($n = 1024, m=32, L = 4, 8$) and
a constant row weight ensemble ($n = 1024, m = 32, r = 100, 200,300)$.
Note that the size of the parity check matrices used in the BP decoder 
is $120 \times 1024$ (redundant random, $L=4$), $1020 \times 1024$ (redundant random, $L=8$),
$32 \times 1024$ (constant row weight), respectively.
It may be observed that the example redundant random ensembles give steeper error curves than those of the example 
 constant row weight ensembles. This difference in decoding performance could be explained from the
the difference in typical stopping distance discussed above.
\hfill\qed
\begin{figure}[htbp]
  \begin{center}
  \includegraphics[scale=0.6]{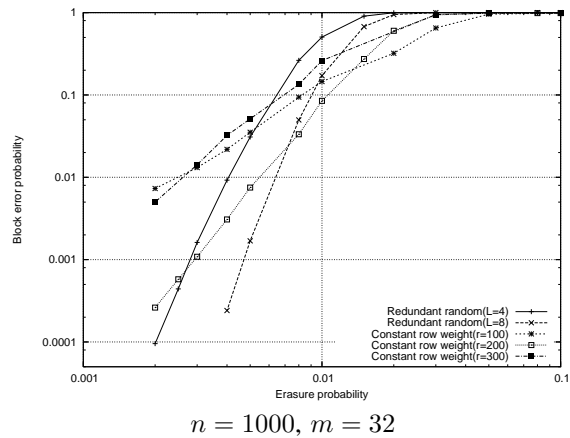} \\
  $n = 1000$, $m = 32$ \\
      \caption{Block error probabilities (BP on BEC) of examples of a redundant random ensemble and constant row 
      weight ensemble.}
    \label{fig:constrandom}
      \end{center}
\end{figure}
\end{example}

\section{Asymptotic growth rate of the average SS weight distributions of redundant random ensembles}
In this section, we will discuss the asymptotic ({\em i.e.}, in the limit as $n$ goes to infinity) behavior of 
the average SS weight distribution.

\subsection{Bounds on asymptotic growth rate}
We will consider the asymptotic behavior of the average SS weight distribution
of the redundant random ensembles.

The asymptotic growth rate $\sigma_\ell(R, \mu)$ is defined by
\begin{equation}
 \sigma_\ell(R, \mu) \defeq \lim_{n \rightarrow \infty}\frac 1 n \log S_{\ell n}^{\Ren_{(1-R)n,n}^{(\mu n)} } ,
\end{equation}
for $0 \le \ell \le 1, 0< \mu \le 1-R$.  The parameter $\mu$ is called {\em normalized extension degree}. It is evident that, from the definition of
redundant extension, the above definition of $\sigma_\ell(R, \mu)$ is well defined 
only if $(1-R)/\mu$ is an integer. 

The next corollary gives a lower bound on $\sigma_\ell(R, \mu)$.
\begin{corollary}
\label{lemma:lower}
The asymptotic growth rate $\sigma_\ell(R, \mu)$ can be lower bounded by
\begin{equation}
\sigma_\ell(R, \mu) \ge 
\left\{
\begin{array}{ll}
H(\ell) - (1-R),  & \ell \le \mu \\
H(\ell), & \ell > \mu.
\end{array}
\right.
\end{equation}
(Proof)
We first consider the case $\ell \le \mu$. From Theorem \ref{upperlowerL}, we have
the following inequality,
\begin{equation}
S_{\ell n}^{\Ren_{(1-R)n,n}^{(\mu n)}}  
\ge
\max\{A'^{(1-R)/\mu},2^{-(1-R)n}\}  {n \choose \ell n},
\end{equation}
where
\begin{equation}
A' \defeq \max\{1 -(2^{\mu n} - 1)2^{- \ell n} \ell n,0 \}.
\end{equation}
It is evident that $1 -(2^{\mu n} - 1)2^{- \ell n} \ell n \rightarrow -\infty$ in the limit as $n$ goes to infinity.
This implies the equality
\begin{equation}
\max\{A'^{(1-R)/\mu},2^{-(1-R)n}\} 
= 2^{-(1-R)n}
\end{equation}
holds for sufficiently large $n$. Upon using this result, we immediately obtain a lower bound, 
\begin{eqnarray}\nonumber
 \sigma_\ell(R, \mu) &=& \lim_{n \rightarrow \infty}\frac 1 n \log S_{\ell n}^{\Ren_{(1-R)n,n}^{(\mu n)} }  \\ \nonumber
 &\ge& \lim_{n \rightarrow \infty}\frac 1 n \log 2^{-(1-R)n} {n \choose \ell n}  \\
  &=& H(\ell) - (1-R).
\end{eqnarray}
We next consider the case $\ell > \mu$. In this case, 
$1 -(2^{\mu n} - 1)2^{- \ell n} \ell n \rightarrow 1$ in the limit as $n$ goes to infinity, and so

\begin{equation}
\max\{A'^{(1-R)/\mu},2^{-(1-R)n}\} 
\rightarrow 1
\end{equation}
in the limit as $n \rightarrow \infty$. Upon using this result, we obtain
\begin{eqnarray} \nonumber
 \sigma_\ell(R, \mu) &=& \lim_{n \rightarrow \infty}\frac 1 n \log S_{\ell n}^{\Ren_{(1-R)n,n}^{(\mu n)} }  \\ \nonumber
 &\ge& \lim_{n \rightarrow \infty}\frac 1 n \log  {n \choose \ell n}  \\
  &=& H(\ell).
\end{eqnarray}
\hfill\qed
\end{corollary}

The next corollary provides an upper bound on $\sigma_\ell(R, \mu)$.
\begin{corollary}
\label{lemma:upper}
The asymptotic growth rate $\sigma_\ell(R, \mu)$ has an upper bound given by
\begin{equation}
 \sigma_\ell(R, \mu)  \le 
 \left\{
 \begin{array}{ll}
 H(\ell) - (1-R) \left(1 - \frac \ell \mu \right), & \ell \le \mu \\
 H(\ell), & \ell > \mu. 
\end{array}
\right.
\end{equation}
(Proof)
The upper bound in Theorem \ref{upperlowerL} can be rewritten in the form
\begin{eqnarray}\nonumber
S_{\ell n}^{\Ren_{(1-R)n,n}^{(\mu n)}}  
\hspace{-3mm}
&\le& \hspace{-3mm}
 \left(\frac{1-\ell n2^{-\ell n}}{2^{\mu n}  \ell n 2^{-\ell n}+1- 2 \ell n2^{-\ell n}}    \right)^{(1-R)/\mu} {n \choose \ell n} \\
&\le& \hspace{-3mm}
\left(\frac{1-\ell n2^{-\ell n}}{2^{(\mu-\ell) n}  \ell n }    \right)^{(1-R)/\mu} {n \choose \ell n}  
\end{eqnarray}
for sufficiently large $n$. The last inequality holds  because $1- 2 \ell n2^{-\ell n}$ is always positive for large $n$.
Thus, the asymptotic growth rate can be bounded from above:
\begin{eqnarray}  \nonumber
\sigma_\ell(R, \mu) 
\hspace{-3mm} &\le& \hspace{-3mm}
H(\ell) + \frac{(1-R)}{\mu}  \lim_{n \rightarrow \infty} \frac 1 n\log\frac{1-\ell n2^{-\ell n}}{2^{(\mu-\ell) n}  \ell n }  \\
\nonumber
&=& \hspace{-3mm}
H(\ell) + \frac{(1-R)}{\mu} \\
\nonumber
&\times& \hspace{-4mm} \lim_{n \rightarrow \infty} \frac 1 n 
\left(\log (1-\ell n2^{-\ell n}) -\log(2^{(\mu-\ell) n}) -\log( \ell n)    \right)  \\
\nonumber
&=& \hspace{-3mm}
H(\ell) + \frac{(1-R)}{\mu}\lim_{n \rightarrow \infty} \frac 1 n 
\left( -(\mu-\ell) n    \right)  \\ \label{bound1}
&=& \hspace{-3mm}
H(\ell) - (1-R)\left(1 - \frac \ell \mu \right) .
\end{eqnarray}
On the other hand, the inequality
\begin{equation}
\frac{1-\ell n2^{-\ell n}}{(2^{\mu n} - 1) \ell n 2^{-\ell n}+1- \ell n2^{-\ell n}}   \le 1
\end{equation}
leads to another (trivial) upper bound on $S_{\ell n}^{\Ren_{(1-R)n,n}^{(\mu n)}} $,
\begin{equation}
S_{\ell n}^{\Ren_{(1-R)n,n}^{(\mu n)}}  \le {n \choose \ell n}.
\end{equation}
The asymptotic form of this upper bound is given by 
\begin{equation} \label{bound2}
\sigma_\ell(R, \mu) \le H(\ell).
\end{equation}
If $\ell < \mu$, the upper bound (\ref{bound1}) gives smaller values  than the trivial bound (\ref{bound2}).
If $\ell > \mu$, the trivial bound (\ref{bound2}) becomes tighter. When $\ell = \mu$, both of
the bounds yield the same value $H(\ell)$.
\hfill \qed
\end{corollary}

Combining the above two corollaries, we can see that
$
\sigma_\ell(R, \mu)  = H(\ell)
$
 for $\ell > \mu$. That is, the upper and lower bounds are asymptotically 
tight when $\ell > \mu$. 

\begin{example}
Figure \ref{fig:upperlower} shows the lower bound (Corollary \ref{lemma:lower}) 
and the upper bound (Corollary \ref{lemma:upper}) 
for the case  $R = 0.5$, $\mu = 0.25$. 
The horizontal axis of Fig.\ref{fig:upperlower} represents the normalized weight $\ell$.
The curve $H(\ell)$ (the asymptotic growth rate of non-extended ensemble 
$\Ren_{(1-R)n,n}$) is also included in Fig.\ref{fig:upperlower} as a reference.
\begin{figure}[htbp]
  \begin{center}
  \includegraphics[scale=0.6]{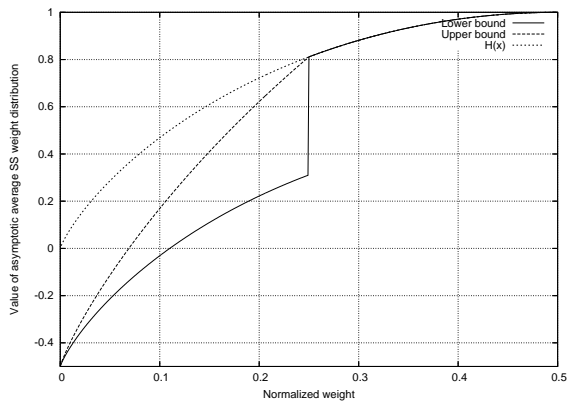} 
      \caption{Upper and lower bounds on the asymptotic growth rate of
         the redundant random ensemble ($R = 0.5$, $\mu = 0.25$).}
    \label{fig:upperlower}
      \end{center}
\end{figure}
\end{example}

\subsection{Critical exponent}
The critical exponent of an ensemble is the normalized weight such that
the asymptotic growth rate changes from negative to positive.
The explicit definition of the critical exponent of a redundant random ensemble is given below:
\begin{definition}[Critical exponent]
The critical exponent of the redundant random ensemble is defined by
\begin{equation}
\alpha(R, \mu)\defeq  \inf \{\ell > 0: \sigma_\ell(R, \mu) \ge 0 \}.
\end{equation}
\hfill\qed
\end{definition}
The following lemma, which gives bounds on the critical exponent, is a 
direct consequence of Corollaries \ref{lemma:lower} and \ref{lemma:upper}.
\begin{lemma}
The following inequality holds
\begin{equation}
\alpha_L(R, \mu) \le \alpha(R,\mu) \le \alpha_U(R, \mu),
\end{equation}
where $\alpha_L(R, \mu)$ is the minimum positive solution of
$
H(\ell) -(1-R) (1 - \ell/\mu) = 0,
$
and $\alpha_U(R, \mu)$ is the minimum positive solution of
$
H(\ell) -(1-R) = 0.
$ \\
(Proof) From Corollaries \ref{lemma:lower} and \ref{lemma:upper}, it is evident that the claim  holds.
\hfill\qed
\end{lemma}
The critical exponent of the bipartite ensemble is given by\cite{Orlitsky},
\begin{equation}
\beta(c,d) \defeq \inf \{\ell > 0: \gamma_\ell(c,d) \ge 0 \}.
\end{equation}

\begin{figure}[htbp]
  \begin{center}
  \includegraphics[scale=0.6]{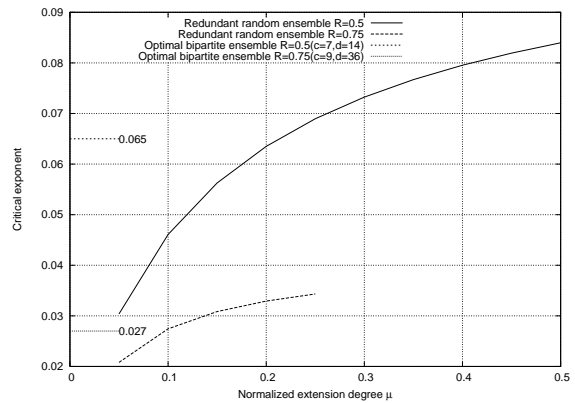} \\
      \caption{Upper bounds on the critical exponent of redundant random ensembles ($R = 0.5, 0.75$).}
    \label{fig:criticalexponent}
      \end{center}
\end{figure}

Figure \ref{fig:criticalexponent} presents 
the lower bound on the critical exponent of the redundant random ensemble with $R = 0.5$ and 
$R = 0.75$. The horizontal axis is the normalized extension degree $\mu$. 
Of course, if $(1-R)/\mu$ is not an integer, the lower bound is not well defined. 
However, for simplicity, the lower bound is plotted as if it were valid in the entire range $0 < \mu \le 1-R$.
We can see that
the exponent increases as $\mu$ increases. Since the parameter $\mu$ can be considered as a 
measure of decoding complexity, the plots in Fig. \ref{fig:criticalexponent} can be regarded as the trade-off curves 
between decoding complexity and the asymptotic decoding performance.

We then compare the critical exponent of the redundant random ensemble and the bipartite 
ensemble with a design rate of 0.5.  It is known that there exists an optimal choice of the
variable node degree $c$ to attain the maximum critical exponent $\beta(c,d)$. The best value is
$
\max_{c \ge 3} \beta(c, 2 c) = 0.065,
$
which is obtained when $c = 7, d = 14$. In other words, no bipartite ensemble with $R = 0.5$ yields 
a critical exponent larger than $0.065$. Note that the plot of the lower bound on the critical exponent of
the redundant random ensemble takes larger values than $0.065$ if $\mu$ is sufficiently large.
This result implies that the asymptotic BP-performance on BEC of a dense matrix can be better than that of 
a sparse matrix. 

In the case of a high code rate (design rate $R=0.75$), we have $\max_{c} \beta(c, 4 c) = 0.027$.
The maximum value is obtained when $c = 9, d=36$. We can observe that, as for the former case, 
the redundant random ensemble with sufficiently large $\mu$ gives larger values.

\begin{example}
Figure \ref{fig:asymSScomp} presents the asymptotic growth rate
of the random ensemble ($R=0.5$), bipartite ensemble $(c=7, d=14)$, 
constant row weight ensemble ($R=0.5, r = 15$) and the redundant random ensemble
($R=0.5, \mu = 0.5$, upper bound). 
The parameters of the bipartite ensemble and the constant row weight 
ensemble are chosen so as to maximize the critical exponent under the constraint $R=0.5$.
In this case, both the bipartite and the constant row weight ensembles have almost the same
maximum critical exponent of 0.065. On the other hand, we have $\alpha_L(0.5, 0.5) =0.083$,
which is larger than the maximum critical exponent of the constant row weight ensemble and
the bipartite ensemble. \hfill\qed
\begin{figure}[htbp]
  \begin{center}
  \includegraphics[scale=0.6]{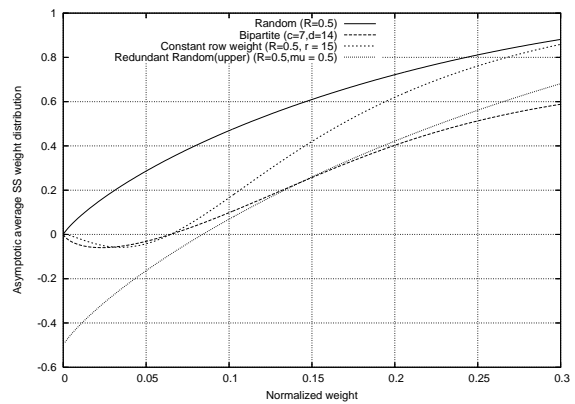} \\
      \caption{Comparison of the asymptotic growth rates: random, bipartite, constant row weight, and
      redundant random ensembles.}
    \label{fig:asymSScomp}
      \end{center}
\end{figure}
\end{example}

\section{Conclusion}
In this paper, the average SS weight distribution and the asymptotic growth rate of 
redundant random ensembles have been analyzed.
The results obtained in the paper describe one aspect of the trade-off between decoding complexity of BP(extension degree) and 
decoding performance.
It is shown that, in some cases,
a dense matrix with linearly dependent rows can yield a better decoding performance
over BEC than a regular LDPC matrix with comparable parameters. In particular, in the high rate regime, a redundant matrix appears to offer promising performance not only 
for BEC, but also for other channels. It is hoped that further research concerning this result can be undertaken.

\section*{Acknowledgment}
 This work was supported by the Ministry of Education, Science, Sports
and Culture, Japan, Grant-in-Aid for Scientific Research on Priority Areas
(Deepening and Expansion of Statistical Informatics) 180790091
and a research grant from SRC (Storage Research Consortium). 

\section*{Appendix}
\subsection*{Proof of the first and second moment of $A_w(G)$}
From a simple counting argument, we obtain $\#\{G \in \Ren_{K,N}: \ve m G = \ve c \} = 2^{(K-1)N}$.
This equation leads to the transformation
\begin{eqnarray} \nonumber
E_{\Ren_{K,N}}[A_w(G)] 
&=& \frac{1}{2^{KN}}   \\ \nonumber 
&\times& \hspace{-5mm} 
\sum_{\ve c \in Z^{(N,w)}}\sum_{\ve m \in \bar F_2^K} \#\{G \in \Ren_{K,N}: \ve m G = \ve c \} \\ \nonumber
&=& \frac{{N \choose w}}{2^{KN}} 
\sum_{\ve m \in \bar F_2^K} \#\{G \in \Ren_{K,N}: \ve m G = \ve c \} \\ \nonumber
&=& \frac{{N \choose w}}{2^{KN}} (2^{K}-1) 2^{(K-1)N} \\
&=&{N \choose w} (2^{K}-1) 2^{-N}.
\end{eqnarray}
We next consider the second moment. The second moment can be written as
\[
E_{\Ren_{K,N}}[(A_w(G))^2] \hspace{5cm}
\]
\begin{eqnarray}
\nonumber
& = &
\sum_{\ve c_1, \ve c_2 \in Z^{(N,w)}} \sum_{\ve m_1, \ve m_2 \in \bar F_2^K}
I[\ve m_1 G = \ve c_1, \ve m_2 G = \ve c_2] \\
\nonumber
&=&
\sum_{\ve c_1, \ve c_2} \sum_{\ve m_1, \ve m_2}I[\ve m_1 \ne \ve m_2] X_1\\
\nonumber
&+&
\sum_{\ve c_1, \ve c_2} \sum_{\ve m_1, \ve m_2}I[\ve m_1 = \ve m_2]I[\ve c_1 \ne \ve c_2] X_2\\
&+&
\sum_{\ve c_1, \ve c_2} \sum_{\ve m_1, \ve m_2}I[\ve m_1 = \ve m_2]I[\ve c_1 = \ve c_2] X_3
\end{eqnarray}
From a combinatorial argument, we have $X_1 = 2^{-2N}, X_2 = 0, X_3 = 2^{-N}$. Finally, we get
\[
E_{\Ren_{K,N}}[(A_w(G))^2] \hspace{5cm}
\]
\vspace{-5mm}
\begin{eqnarray}
\nonumber
&=&
{N \choose w}^2 \left((2^K-1)^2- (2^K-1) \right)2^{-2N}  \\ \nonumber
&+& {N \choose w} (2^K -1 ) 2^{-N}\\ \nonumber
&=& \left(E_{\Ren_{K,N}}[A_w(G)] \right)^2  \\
&+& E_{\Ren_{K,N}}[A_w(G)]  \left(1 - {N \choose w}2^{-N} \right).
\end{eqnarray}

\subsection*{Redundant constant row weight ensemble: $L=2$}
As in the case of the redundant random ensemble, 
it is possible to write down a simple formula for
the redundant constant row weight ensembles if $L = 2$. 
\begin{lemma}\label{lemmapp}
The average SS weight distribution of the redundant constant row weight ensemble with
parameters $n,m (1 \le m < n), r (r \ge 1)$ and $L=2$ is given by 
\begin{equation}
S_w^{\Cen_{n,m,r}^{(2)}} 
=\frac{{n \choose w}}{{n \choose r}^m}  \left( \left( {n \choose r} - w {n-w \choose r -1}\right )^2 
- V \right)^{m/2},
\end{equation}
where $V$ is given by
\begin{equation}
V = 2 \sum_{\gamma = 2}^{\min\{w-1,r-1\}}(w - \gamma) {w \choose \gamma}{n-w \choose r - \gamma - 1} {n-w \choose r - \gamma }.
\end{equation}
(Proof) The proof of the theorem is almost same as the proof of Theorem \ref{thave2}.
Let 
\begin{eqnarray}\nonumber
U \defeq \# \{(\ve h_1,\ve h_2) \in Z^{(n,r)} \times Z_{(n,r)}: \ve h_1 \circ \ve x_w \ne 1,  \\
\ve h_2 \circ \ve x_w \ne 1,
(\ve h_1 \oplus \ve h_2) \circ \ve x \ne 1   \} .
\end{eqnarray}
and
\begin{eqnarray} \nonumber
V \defeq
\# \{(\ve h_1, \ve h_2) \in Z^{(n,r)} \times Z^{(n,r)}: \ve h_1 \circ \ve x \ne 1,  \\
\ve h_2 \circ \ve x \ne 1, (\ve h_1 \oplus \ve h_2) \circ \ve x = 1 \}
\end{eqnarray}
A simple combinatorial argument similar to that used in the proof of Theorem \ref{thave2} leads to a simple formula 
for $V$, 
\begin{equation}
V = 2 \sum_{\gamma = 2}^{\min\{w-1,r-1\}}(w - \gamma) {w \choose \gamma}{n-w \choose r - \gamma - 1} {n-w \choose r - \gamma }
\end{equation}
Upon using the relation
\begin{equation}
U = \# \{(\ve h_1, \ve h_2) \in Z^{(n,r)} \times Z^{(n,r)}: \ve h_1 \circ \ve x \ne 1, \ve h_2 \circ \ve x \ne 1\} - V,
\end{equation}
we have 
\begin{equation}
U
= \left( {n \choose r} - w {n-w \choose r -1}\right )^2 - V
\end{equation}
for $0 \le w \le n, r \ge 2$. The average SS weight distribution is therefore given by
\begin{eqnarray} \nonumber
S_w^{\Cen_{n,m,r}^{(2)}} 
&=& \frac{1}{\# \Cen_{n,m,r}^{(2)}}\sum_{\ve x \in Z^{(n,w)}}   \#\{H \in \Cen_{n,m,r}^{(2)}: q_H(\ve x) = 0 \} \\ \nonumber
&=& \frac{{n \choose w}}{{n \choose r}^m}    \#\{H \in \Cen_{n,m,r}^{(2)}: q_H(\ve x_w) = 0 \} \\
&=& \frac{{n \choose w}}{{n \choose r}^m}   U^{m/2},
\end{eqnarray}
where the last inequality coincides with the claim of the lemma.
\hfill\qed
\end{lemma}

\end{document}